\documentclass[aps,prb,twocolumn,showpacs]{revtex4}

\usepackage{amsmath}
\usepackage{amsfonts}
\usepackage{amssymb}
\usepackage{graphicx}
\usepackage{dsfont}
\usepackage{sidecap}

\newcommand{\abs}[1]{\ensuremath{\left\vert#1\right\vert}}

\newcommand{\heaviside}[1]{\ensuremath{\theta\left[#1\right]}}
\renewcommand{\Re}[1]{\ensuremath{\operatorname{Re} \left\{#1\right\} } }
\renewcommand{\Im}[1]{\ensuremath{\operatorname{Im} \left\{#1\right\} } }

\renewcommand{\vec}[1]{\boldsymbol{#1} }

\begin{document}

\title{Plasmons in spin-orbit coupled two-dimensional hole gas systems}
\author{Andreas Scholz}
\email[To whom correspondence should be addressed. Electronic
address:]{andreas.scholz@physik.uni-regensburg.de}
\author{Tobias Dollinger}
\author{Paul Wenk}
\author{Klaus Richter}
\author{John Schliemann}
\affiliation{Institute for Theoretical Physics, 
University of Regensburg, D-93040 Regensburg, Germany}
\date{\today}

\begin{abstract}
We study the dynamical dielectric function of a two-dimensional hole gas,
exemplified on [001] GaAs and InAs quantum wells,
within the Luttinger model extended to the two lowest subbands including bulk and 
structure inversion asymmetric terms.
The plasmon dispersion shows a pronounced anisotropy for GaAs- and InAs-based 
systems.
In GaAs this leads to a suppression of plasmons due to Landau damping in some 
orientations.
Due to the large Rashba contribution in InAs, the lifetime of plasmons can be 
controlled
by changing the electric field. This effect is potentially useful in plasmon 
field effect transistors as previously proposed for electron gases.
\end{abstract}


\maketitle

\section{Introduction}
\label{sect:Introduction}
Spin-orbit coupling (SOC), as one of the most important consequences of Dirac's theory for particles
in solid state physics, is widely believed to play an important role for establishing electronic devices such as spintronic transistors.\cite{Zutic_2004, Fabian_2007, Wu_2010}
By varying the single-particle properties of materials with sufficiently large spin-orbit coupling,
typical response quantities such as the conductivity or the dielectric function can be controlled in an efficient way.\cite{Li_2008, Badalyan_2009, Kyrychenko_2009}
The latter is not only necessary in understanding screening of extrinsic charged impurities, which in turn is
important for transport, but also because of the existence of collective charge excitations known as plasmons.\cite{Guil, Fetter}

Many analytical and numerical studies have been made in the last years regarding the dielectric properties of 
electron\cite{Li_2008, Badalyan_2009, Badalyan_2010, Ullrich_2003, Pletyukhov_2007} and hole gas systems\cite{Schliemann_2010, Kernreiter_2010, Kyrychenko_2011, Cheng_2001}
or promising materials like graphene.\cite{Wunsch, Sarma, Pyat, Stauber_2010, Scholz_2011, Scholz_2012, Wang_2007_2}
As has been shown in recent works, large analytical progress could be made in
a two-dimensional electron gas (2DEG) including the effect of an asymmetric confinement, the Rashba SOC, and the contribution due to bulk inversion asymmetry (BIA), the so called Dresselhaus SOC,\cite{Pletyukhov_2007, Badalyan_2009, Badalyan_2010}
or in graphene including several types of spin-orbit interactions (SOIs).\cite{Pyat, Scholz_2011, Scholz_2012}
Due to the more complicated nature of the valence bands, the calculation of the dielectric function in hole gas systems
turns out to be a formidable task even without SOIs.
While for the three-dimensional case the free polarizability has been obtained for arbitrary frequencies and wave vectors
within the axial model,\cite{Schliemann_2010} the solution for the two-dimensional case is not known so far.

In this work, we use Luttinger's four-band model\cite{Luttinger_1956} extended to the two lowest subbands
including the lowest order Dresselhaus and Rashba contribution and discuss the dielectric functions of GaAs- and InAs-based quantum wells.
As we will show, the plasmon mode exhibits a pronounced anisotropy being much stronger in GaAs than in InAs.
By applying an electric field perpendicular to the 2D hole gas in InAs, giving rise to a tunable spin-splitting of the bands due to the
Rashba coupling, it is possible to modulate the damping rate of the plasmons.

This paper is organized as follows.
In Sec.~\ref{sect:Model} we introduce Luttinger's model including SOIs of the Dresselhaus and Rashba type
and summarize the formalism of random phase approximation.\cite{Guil, Fetter}
In Sec.~\ref{sect:DOS} the energy spectrum and the density of states of a GaAs and InAs quantum well
is discussed.
In Sec.~\ref{sect:Plasmons} the anisotropy of the plasmon spectrum and the influence of a finite electric field on the lifetime of
charge excitations is studied and possible applications are described.
Finally, in Sec.~\ref{sect:Conclusions} we give a brief summary and a short outlook.

\section{The model}
\label{sect:Model}

\subsection{Hamiltonian}
We start with the basic part of Luttinger's Hamiltonian\cite{Luttinger_1956} for III-V semiconductors in three dimensions (setting $\hbar = 1$ throughout this work),
\begin{align}
\hat H_0 =& \frac{1}{2m_0} \left[ \gamma_1 \left(\hat k_x^2 + \hat k_y^2 + \hat k_z^2\right)
- 2\gamma_2 \left(\hat k_x^2\left(\hat J_x^2 - \frac 13 \vec{\hat J}^2 \right) + \right. \right. \notag \\ 
& \qquad \text{c.p.} \bigg) - 4\gamma_3 \left( \left\{\hat J_x, \hat J_y \right\} \left\{\hat k_x, \hat k_y \right\}
+ \text{c.p.} \right)
\bigg] \, , \label{Hamiltonian_H0}
\end{align}
where c.p.~stands for cyclic permutation of variables, $\hat J_i$ are the usual spin-3/2 matrices, and $m_0$ is the bare electron mass.

\begin{table*}[t]
\caption{Band structure and spin-orbit coupling parameters of GaAs and InAs. Taken from Ref. \cite{Winkler_Book}.}
\begin{minipage}{10cm}
    \begin{tabular}{ | l | l | l | l | l | l | l | l | l |}
    \hline
     & $\gamma_1$ & $\gamma_2$ & $\gamma_3$ & $C_k$ (eV\AA)& $b_{41}^{8v8v}$ (eV\AA$^3$)& $r_{41}^{8v8v}$ (e\AA$^2$) & $\epsilon_r$ & $E_0(20$nm$)$ (meV) \\ \hline
    GaAs & 6.85 & 2.10 & 2.90 & -0.0034 & -81.93 & -14.62 & 12.5 & 6.4 \\ \hline
    InAs & 20.40 & 8.30 & 9.10 & -0.0112 & -50.18 & -159.9 & 14.6 & 19.3 \\ \hline
    \end{tabular}
\end{minipage}
\label{TABLE_Parameters}
\end{table*}

Often $\hat H_0$ is simplified by setting $\gamma_2 = \gamma_3 = \bar\gamma =  \left(2\gamma_2+3\gamma_3\right) / 5$.
The advantage of this so-called \textit{axial} (in two spatial dimensions) and \textit{spherical} (in three dimensions) approximation
is the relative simplicity of the energy spectrum.
In three dimensions the subsequent twofold spin-degenerated bands read
\begin{align}
E_{h/l}^{sp} (\vec k, k_z) = \frac {\vec k^2 + k_z^2}{2m_{h/l}} \; , \label{Energy_axial_3D}
\end{align}
where $m_{h/l} = m_0 / (\gamma_1 \mp 2\bar\gamma)$ is the heavy and light hole mass
and $\vec k = (k_x, k_y)$ the wave vector in the $x$-$y$ plane.
While these bands resemble that of an ordinary electron gas,
the eigenstates are clearly more complicated\cite{Schliemann_2010} due to their non trivial spin structure.

In addition to the Hamiltonian in Eq. (\ref{Hamiltonian_H0})
we take into account SOIs of the Dresselhaus type 
up to the third order in ${\bf k}$ (which indeed is the lowest order to describe the effect due to BIA), \cite{Dresselhaus_1955}
\begin{align}
&\hat H_D = -\frac {2C_k}{\sqrt 3} \left[\hat k_x \left\{\hat J_x, \hat J_y^2 - \hat J_z^2\right\} + \text{c.p.} \right] \notag \\
& \qquad \quad - b_{41}^{8v8v} \left[ \left\{\hat k_x, \hat k_y^2 - \hat k_z^2\right\} \hat J_x + \text{c.p.} \right] , \label{Hamiltonian_HD}
\end{align}
and the Rashba\cite{Rashba_1984} contribution,
\begin{align}
&\hat H_R = -r_{41}^{8v8v} \left[ \left(\hat k_y E_z - \hat k_z E_y \right) \hat J_x + \text{c.p.} \right] , \label{Hamiltonian_HR}
\end{align}
derived perturbatively by using the Loewdin approximation.\cite{Winkler_Book}
While the former is always present provided bulk inversion symmetry is broken, as is the case for zinc blende structures,
the latter arises, e.g., due to a suitable external electric field breaking structure inversion symmetry.
Notice that the signs of the Hamiltonians $H_0$, $H_D$, and $H_R$ are inverted such that all energies are positive.

We now add a spatial confinement in the [001] direction and express the Hamiltonian in the basis of the subband functions
$\varphi_n (z) = \sqrt{2/d} \sin(n\pi z/d)$ ($n \in \mathbb{N}$),
with $d$ being the width of the quantum well.
It is worth noticing that both the contribution due to the Dresselhaus SOI as well as the one due to the Rashba SOI in the 
two-dimensional hole gas (2DHG) depend on the growth direction of the crystal.\cite{Winkler_Book}
However, we left out the terms in Eq.~(\ref{Hamiltonian_HR}) which result in a dependence on the crystallographic orientation of both the applied external field and the momentum ${\bf k}$ since they can be considered as much smaller than $r_{41}^{8v8v}$.\cite{Winkler_Book}\\
The new matrix contains blocks of four-dimensional Hamiltonians, each given by the projection
$\left\langle \varphi_n \vert \hat H_0 + \hat H_D + \hat H_R \vert \varphi_m  \right\rangle$.
For $\hat k_z$ this leads to 
\begin{align*}
\left\langle \varphi_n  \Big\vert \hat k_z \Big\vert \varphi_m  \right\rangle
= &\frac{4imn}{d\left(m^2-n^2\right)} \sum_{i=0}^\infty \delta_{\abs{n-m},2i+1}
\end{align*}
and
\begin{align*}
\left\langle \varphi_n  \Big\vert \hat k_z^2 \Big\vert \varphi_m  \right\rangle
= \frac{n^2\pi^2}{d^2} \delta_{n,m} .
\end{align*}

Often the off-diagonal contributions ($n\neq m$) are neglected. In this case the mode $n$ is a good quantum number
and $\varphi_n (z)$ is the exact eigenfunction in the $z$ direction.
However, this approximation is only justifiable if $\gamma_3 \ll \gamma_1$ which is not the case for GaAs
and InAs; see Table \ref{TABLE_Parameters}.

\subsection{Dielectric function}
\begin{figure}[b]
\includegraphics[width=.8 \linewidth]{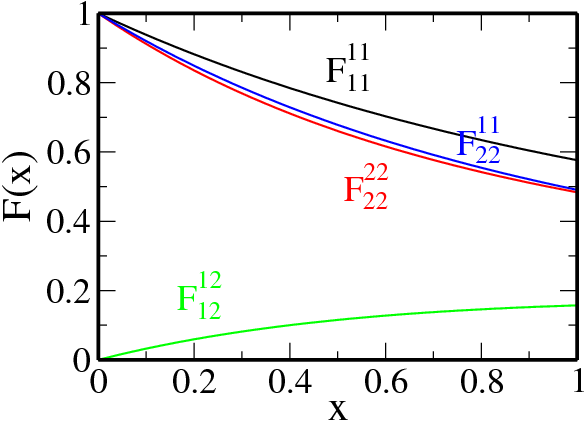}
\caption{(Color online) Independent finite form factors $F^{n_1n_3}_{n_2n_4}$ as defined in Eq. (\ref{DEF_Formfactor}).
}
\label{Formfactor}
\end{figure}

The charge susceptibility of the non interacting system can be expressed as\cite{Guil, Fetter}
\begin{align}
\chi_{0} (\vec q,\omega) = & \sum_{\lambda_1,\lambda_2} \int \frac {d^2k}{(2\pi)^2}
\frac {f\left[E_{\lambda_1}(\vec k)\right] - f\left[E_{\lambda_2}(\vec{k+q})\right]} {\omega - E_{\lambda_2}\left(\vec k+\vec q\right) + E_{\lambda_1}\left(\vec k\right) + i0} \times \notag \\
& \times \left( \sum_{ {n_1,n_2, \atop n_3,n_4}} \left\langle\chi_{\lambda_1}^{n_1}\left(\vec k\right) \bigg| \chi_{\lambda_2}^{n_3}\left(\vec k+\vec q\right) \right\rangle \right. \notag \\
& \times \left. \left\langle\chi_{\lambda_2}^{n_2}\left(\vec k+\vec q\right) \bigg| \chi_{\lambda_1}^{n_4}\left(\vec k\right) \right\rangle
F^{n_1n_3}_{n_2n_4}(q) \right)
, \label{DEF_suscept}
\end{align}
where the $E_{\lambda}$ are the eigenenergies and $\vert\chi_{\lambda}^{n_i}\rangle$
the \mbox{$n_i$-th} components of the eigenstates ($i=1,...,4)$.
The above expression contains a summation over all band indices $\lambda_i$ and modes $n_i$.
The form factor\cite{Guil, Badalyan_2008} $F^{n_1n_3}_{n_2n_4}(q)$, defined as
\begin{align}
F^{n_1n_3}_{n_2n_4}(q) = & \int_{0}^d dz \int_{0}^d dz'
\varphi_{n_1}(z) \varphi_{n_3}(z) \times \notag \\
& \qquad \times \varphi_{n_2}(z') \varphi_{n_4}(z') e^{-\abs{z-z'}q} , \label{DEF_Formfactor}
\end{align}
arises due to the spatial confinement in the growth direction.
For the two lowest subbands only six independent form factors remain due to symmetry, i.e., 
$F^{11}_{11}$, $F^{22}_{22}$, $F^{12}_{12}$, $F^{11}_{22}$, $F^{11}_{12}$, and $F^{12}_{22}$. 
While the latter two vanish, the four finite contributions,
which have been calculated numerically, are shown in Fig.~\ref{Formfactor}.

The dependence on temperature and chemical potential is contained in the Fermi distribution function $f(E)$.
In this work, we assume zero temperature such that $f(E) = \heaviside{\mu - E}$, with $\mu$ being the chemical potential.
The general relation $\chi_0(q,-\omega) = \left[ \chi_0(q,\omega) \right]^*$,
valid for all response quantities, insures that we can restrict ourselves to positive frequencies $\omega$.

Having found the non interacting charge response, the dielectric function in random phase approximation (RPA),
i.e., the system's response to a scalar probe potential, immediately follows from\cite{Guil, Fetter}
\begin{align*}
\varepsilon(\vec q,\omega) = 1 - V(q) \chi_0 (\vec q,\omega) .
\end{align*}
$V(q) = e^2/(2\epsilon_0 \epsilon_r q)$ is the Fourier transform of the Coulomb potential in two dimensions,
$\epsilon_0$ the vacuum permittivity, and $\epsilon_r$ the background dielectric constant.
Note that as the Hamiltonian is anisotropic, the dielectric function will be a function of both
$q = \abs{\vec q}$ and the polar angle given by $\tan\phi_q=q_y/q_x$.

\begin{figure}[b]
\includegraphics[scale=0.34]{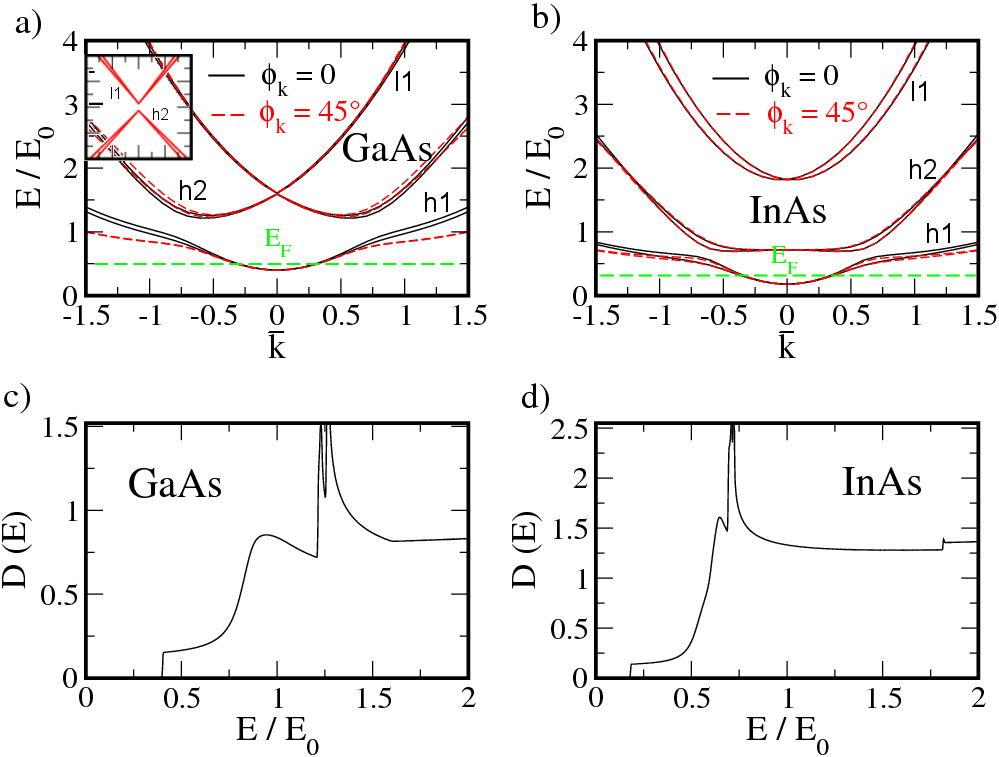}
\caption{(Color online) Top: Energy spectrum of (a) GaAs and (b) InAs for two different orientations
$\phi_k = 0^\circ$ (solid black line) and $\phi_k = 45^\circ$ (dashed red), with $\tan{\phi_k} = k_y/k_x$.
The dashed green line indicates the Fermi energy $E_F$ corresponding to a carrier density of $n=5\cdot 10^{14}$ m$^{-2}$.
The inset in (a) shows the energies $E_{h2}$ and $E_{l1}$ around the $\Gamma$ point.
Bottom: Density of states of (c) GaAs and (d) InAs.
In all cases, the electric field is set to $E_z = 1.5\cdot 10^7$ V/m.
}
\label{DOS}
\end{figure}

\section{Dispersion and density of states of a quantum well}
\label{sect:DOS}
Before we start with the discussion of the dielectric function,
we have a closer look on the energy spectrum and the density of states (DOS)
\begin{align}
D(E) = \sum_{\lambda} \sum_{\vec k} \delta \left[ E - E_\lambda(\vec k) \right] . \label{Def_DOS}
\end{align}
The latter is related to the imaginary part of the polarizability through the Dirac identity,
$\Im{1 / (x \pm i0)} = \mp \pi \delta(x)$, applied to Eq.~(\ref{DEF_suscept}).

In what follows, we choose parameters according to a GaAs and InAs quantum well grown in the [001]
direction\cite{Winkler_Book} with a thickness of $d=20$ nm, an electric field of order $10^7$ V/m,
and a hole concentration of $n = 5 \cdot 10^{14}$ m$^{-2}$.\cite{Schueller_1994}
The band structure parameters can be found in Table \ref{TABLE_Parameters}.
We restrict ourselves to the two lowest subbands ($l=1,2$) and neglect higher ones, as well as electronic and split off bands. 
Our numerical inspection shows that the dominant contributions in the dielectric function arise from the intraband
and interband transitions with final states in the ground state light and heavy hole and in the first excited heavy hole band, respectively.
Already the inclusion of the $l=2$ light hole states does not lead to significant changes in the plasmon spectrum.
Hence the influence of energetically higher bands such as the $l=3,4,...$ subbands or the electronic and split-off bands will be even smaller.

For convenience, we furthermore introduce the energy scale $E_0 = \pi^2\gamma_1/(2m_0d^2)$
and use the notation that all wave vectors with a \textit{bar} symbol have to be understood
as dimensionless quantities measured in units of $\pi/d$; e.g., $\bar q = qd/\pi$.

\begin{figure}[t]
\includegraphics[width=\linewidth]{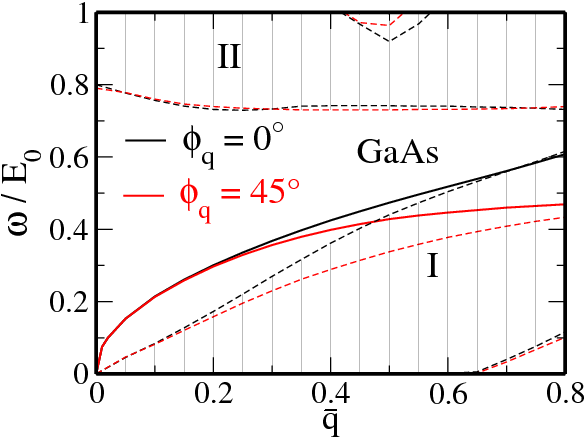}
\caption{(Color online) The solid lines show the plasmon dispersion of GaAs for an angle orientation of $\phi_q = 0^\circ$ (black) and $\phi_q = 45^\circ$ (red). The dashed lines display the boundaries of the single-particle continuum for $\phi_q = 0^\circ$ (black) and $\phi_q = 45^\circ$ (red). The electric field is set to $E_z = 0$ and the hole density to $n=5\cdot 10^{14}$ m$^{-2}$ ($E_F = 0.5 \cdot E_0$).}
\label{Plasmons_GaAs}
\end{figure}

\subsection{GaAs}
In Fig.~\ref{DOS}(a) the six lowest valence bands of GaAs are shown for an electric field of $E_z = 1.5 \cdot 10^7$ V/m.
The solid black (dashed red) curve shows an in-plane momentum angle orientation of $\phi_k = 0^\circ$ ($45^\circ$),
where $\tan\phi_k = k_y / k_x$.

Neglecting the off-diagonal contributions, the heavy and light hole energies of the $n$th subband right at the $\Gamma$-point can be approximated by $E_{h/l, n}(0)=n^2 \left( 1 \mp 2\gamma_2/\gamma_1 \right) E_0$.
The two lowest bands, $E_{h1}(0) = 0.38 \cdot E_0$, are well separated from the others.
However, the four next states, $E_{h2}$ and $E_{l1}$, are very close to each other,
though they do not touch which can be seen from the inset in Fig.~\ref{DOS}(a).
Thus, in GaAs it is not sufficient to restrict to the ground state energies ($n=1$) but one also needs
to take into account the first excited subband.\cite{Cheng_2001}

For large enough momenta, $\bar k > 0.5$, the anisotropy of the heavy hole bands
is clearly larger than that of the light holes.
Additionally, the Dresselhaus contribution leads to a spin splitting of the bands.
As the Rashba part in GaAs is virtually negligible for realistic fields,\cite{Winkler_Book}
only minor changes occur in the spectrum once an electric field is turned on.

The DOS obtained from Eq.~(\ref{Def_DOS}) is shown in Fig.~\ref{DOS}(c).
First of all, due to quantization in the $z$ direction, the hole DOS is zero for energies smaller than $E_{h1}(0)$.
Second, as the energy spectrum shows singular points at $E_{vH} \approx 1.25 \cdot E_0$, see Fig.~\ref{DOS}(a),
we observe two nearby Van Hove singularities around $E_{vH}$.
Without SOIs, spin degeneracy is recovered and both singularities merge.
Finally, for large enough energies, $E > 2 \cdot E_0$, the DOS remains roughly constant.

\subsection{InAs}
The numerically calculated energy spectrum of InAs, including a finite electric field of $E_z = 1.5 \cdot 10^7$ V/m, is shown in Fig.~\ref{DOS}(b).
Compared to GaAs, the anisotropy in the spectrum of InAs is much weaker.
As the anisotropic terms are proportional to $\gamma_2 - \gamma_3$,
the reason for this is the smaller relative difference of $\gamma_2$ and $\gamma_3$, roughly being $10$ percent in InAs while in GaAs it is about $30$ percent; see Table \ref{TABLE_Parameters}.
Unlike GaAs, the Rashba contribution now dominates over the Dresselhaus part.
While for zero electric field the band spin-splitting is small,
a noticeable spin splitting, as shown in Fig.~\ref{DOS}(b), is mainly caused by the Rashba part.
Another important difference between GaAs and InAs is that the $\Gamma$-point energy of the first excited heavy hole state, i.e.,
$E_{h2} = 0.74 \cdot E_0$, is much smaller than that of the ground state light hole band, $E_{l1} = 1.81 \cdot E_0$.
Hence the $n=2$ subband cannot be neglected.

The DOS of InAs shows features similar to those of GaAs; see Fig.~\ref{DOS}(d).
That is, the hole DOS is zero for energies $E < E_{h1} = 0.17 \cdot E_0$.
The step at $E_{h1}$ describes the sudden occupation of the two lowest bands.
Furthermore, singular points in the derivative of the energy dispersion at $E\approx 0.70 \cdot E_0$ and $E \approx E_{l1}$
give rise to characteristic spin-split Van Hove singularities.

\section{Collective charge excitations}
\label{sect:Plasmons}
We now continue with the discussion of collective charge excitations.
Plasmons are defined as zeros of the dielectric function,\cite{Guil, Fetter}
\begin{align}
\varepsilon (\vec q, \omega_q) = 0 , \label{Def_Plasmons}
\end{align}
where $\omega_q$ is the plasmon energy.
For small damping rate, Eq. (\ref{Def_Plasmons}) can be approximated by
\begin{align}
\Re{\varepsilon (\vec q, \omega_q) } = 0 . \label{Def_weakly_damped_Plasmons} 
\end{align}

\begin{figure}[b]
\includegraphics[width=\linewidth]{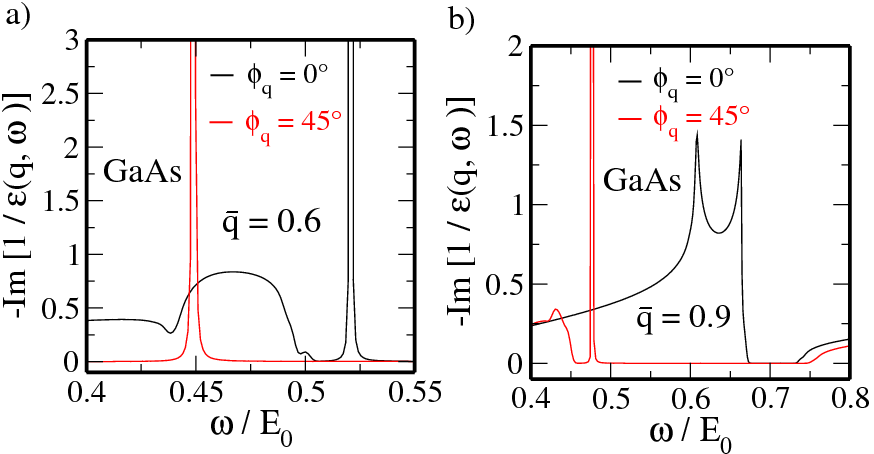}
\caption{(Color online) Energy loss function of GaAs for two different angle orientations, $\phi_q = 0^\circ$ (black) and $\phi_q = 45^\circ$ (red),
and momenta (a) $\bar q = 0.6$ and (b) $\bar q = 0.9$.
The electric field is set to $E_z = 0$ and the charge carrier concentration to $n=5\cdot 10^{14}$ m$^{-2}$ ($E_F = 0.5 \cdot E_0$).
}
\label{lossfunction_GaAs}
\end{figure}

\begin{figure}[tb]
\includegraphics[width=\linewidth]{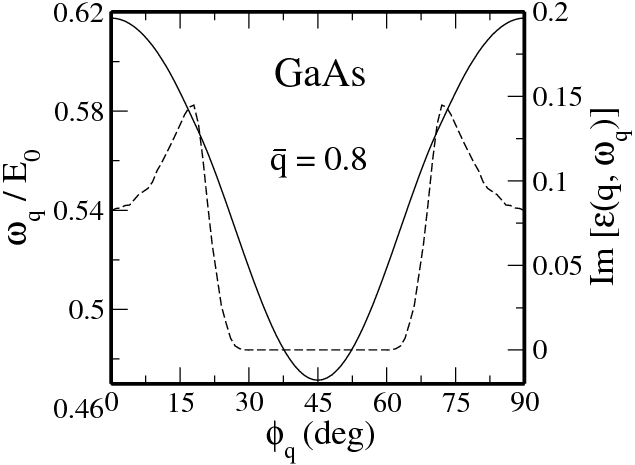}
\caption{(Color online) Angle dependence of the plasmon energy (solid line) and imaginary part of the dielectric function (dashed line) for GaAs with fixed $\bar q = 0.8$.
The electric field is set to $E_z = 0$ and the charge carrier concentration to $n=5\cdot 10^{14}$ m$^{-2}$ ($E_F = 0.5 \cdot E_0$).
}
\label{Plasmons_angle_dependence}
\end{figure}

In principle, one can solve Eq.~(\ref{Def_weakly_damped_Plasmons}) for arbitrary wave vectors.
However, these mathematical solutions correspond only at sufficiently low 
damping to a physical resonance. A practical way to clarify this is to 
study the energy loss function, $\Im{1 / \varepsilon(\vec q,\omega+i0)}$, which
is relevant in experiments.

\subsection{GaAs}
In a recent work\cite{Kyrychenko_2011} using a multiband $\vec k \cdot \vec p$ model it has been demonstrated that plasmons in GaAs-based three-dimensional hole gas structures behave quite differently compared to their electronic counterpart.
In particular, in three dimensions the plasmon energy always lies in the single-particle continuum, even for small momenta.
Contrary to electronic systems, plasmons are therefore always Landau damped.

\begin{figure}[tb]
\includegraphics[scale = 0.45]{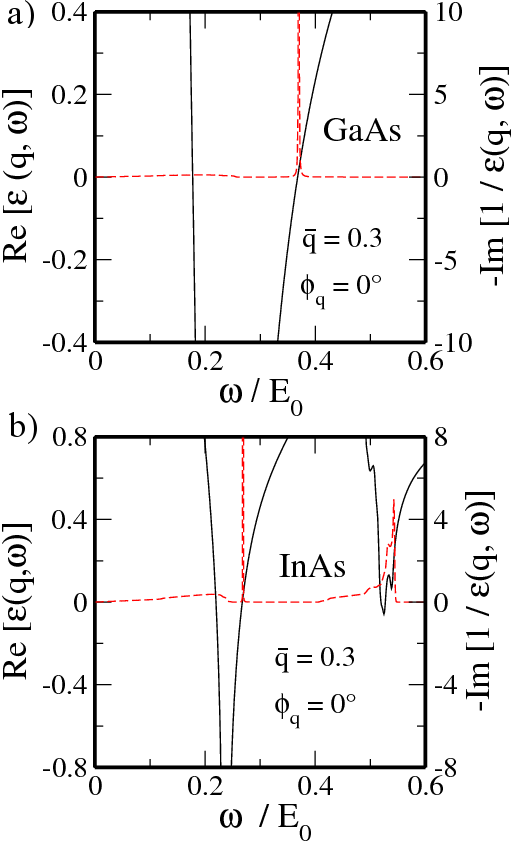}
\caption{(Color online) (a) Real part of the dielectric function (solid black) and imaginary part of the energy loss function (dashed red)
of GaAs. (b) The same for InAs.
Parameters: $\bar q = 0.3$, $\phi_q = 0^\circ$, $E_z = 0$, and $n=5\cdot 10^{14}$ m$^{-2}$.
}
\label{Plasmons_2nd_zero}
\end{figure}

In Fig.~\ref{Plasmons_GaAs}, we show the numerically calculated plasmon dispersion of a GaAs quantum well
obtained from Eq.~(\ref{Def_weakly_damped_Plasmons})
for two different polar angles $\phi_q = 0^\circ$ (solid black line) and $\phi_q=45^\circ$ (solid red).
As mentioned above, the hole density is set to $ n = 5 \cdot 10^{14}$ m$^{-2}$ ($E_F = 0.5 \cdot E_0$);
i.e., the two lowest bands are occupied.
The black and red dashed lines indicate the boundaries of the single-particle continuum (SPC) in the respective orientation
in order to see when plasmons acquire a finite lifetime.
Region I is thereby caused by intraband transitions between the lowest lying valence bands ($E_{h1}$), while
the interband continuum II is due to transitions with final bands $E_{h2}$ and $E_{l1}$, respectively.

As can be seen from Fig.~\ref{Plasmons_GaAs}, the lower part of the SPC (region I) and the plasmon spectrum turn out to be clearly anisotropic.
Furthermore, the $\phi_q = 0^\circ$ mode becomes damped at smaller wave vectors compared to the $\phi_q = 45^\circ$ solution.
To see this, the energy loss function $\Im{1/\varepsilon(\vec q,\omega+i0)}$ is plotted in Fig.~\ref{lossfunction_GaAs}.
For $\bar q = 0.6$, see Fig.~\ref{lossfunction_GaAs}(a), in both cases distinct peaks occur in the loss function indicating truly coherent modes.
For larger wave vector, $\bar q = 0.9$ in Fig.~\ref{lossfunction_GaAs}(b), the peak is smeared out in the $\phi_q = 0^\circ$ direction, i.e.,
the mode is damped, while for $\phi_q = 45^\circ$ it remains $\delta$-like.

In Fig.~\ref{Plasmons_angle_dependence}, the angle dependence of the plasmon energy and imaginary part of the dielectric function
are shown for fixed $\bar q = 0.8$.
The energy (solid line) is redshifted for larger angles, up to a minimal value at $\phi_q = 45^\circ$.
The relative difference between the $\phi_q = 0^\circ$ and $\phi_q = 45^\circ$ energies is roughly $30$ percent and thus of order of
the difference of the hole gas parameters $\gamma_2$ and $\gamma_3$.
The second (dashed) line in the plot shows the imaginary part of the dielectric function
being proportional to the quasi particle lifetime.
For $\phi_q \le 29^\circ$ the damping rate is finite with a maximal value at around $\phi_q = 18^\circ$.
For $30^\circ \le \phi_q \le 45^\circ$ the imaginary part vanishes and the peak
in the loss function becomes sharp.

It should also be emphasized that for a given momentum, Eq.~(\ref{Def_weakly_damped_Plasmons}) leads to an additional
solution with smaller energy.
However, this solution does not fulfill the original condition of Eq.~(\ref{Def_Plasmons}) as the imaginary part of the
dielectric function is not negligible.
This is exemplarily illustrated in Fig.~\ref{Plasmons_2nd_zero}(a) where the real part of the dielectric function (solid black) 
and the energy loss function (dashed red) are shown.
From the latter one can see that only one resonance occurs and thus the additionally found solution is physically irrelevant.

\begin{figure}[tb]
\includegraphics[width=\linewidth]{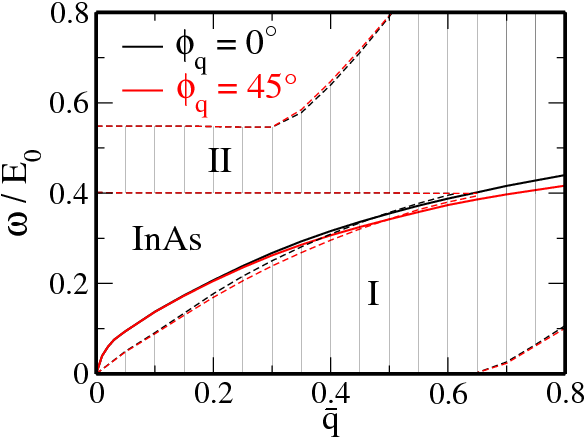}
\caption{(Color online) The solid lines show the plasmon dispersion of InAs
for an angle orientation of $\phi_q = 0^\circ$ (black) and $\phi_q = 45^\circ$ (red).
The dashed lines display the boundaries of the single-particle continuum for
$\phi_q = 0^\circ$ (black) and $\phi_q = 45^\circ$ (red).
The electric field is set to $E_z = 0$ and the hole density to $n=5\cdot 10^{14}$ m$^{-2}$ ($E_F = 0.3 \cdot E_0$).
}
\label{Plasmons_InAs}
\end{figure}

\subsection{InAs}
So far we have seen that plasmons in GaAs-based two-dimensional hole systems show a pronounced anisotropy
as manifested, e.g., in the plasmon spectrum and damping rate.
Unfortunately, as the Rashba contribution in GaAs is virtually negligible, see Table \ref{TABLE_Parameters},
there is no direct control of this feature, e.g., by turning on/off the anisotropy dynamically by varying an electric field.
This control, however, is necessary for possible applications, e.g., as a plasmon filter or a 
plasmon field-effect transistor, as already proposed for 2DEGs.\cite{Badalyan_2009, Li_2008}
On the other hand, as we know that in InAs the Rashba parameter and its influence on the energy spectrum is large,
we expect SOC to give the possibility to modulate the plasmon spectrum as well.
In principle, one might also use materials with an even larger Rashba coupling constant than that of InAs, e.g., InSb.
However, the reason we focus on the former is twofold.
First of all, in experiments InAs is more popular than InSb
and, second, the four-band Luttinger Hamiltonian used in this work is not sufficient to
describe InSb where the more advanced $8\times8$ Kane model including 
electronic conduction bands and split off bands needs to be used.\cite{Kane_1957}

In Fig.~\ref{Plasmons_InAs} the plasmon energy and the SPC of InAs without an electric field is shown for $\phi_q = 0^\circ$ (solid black)
and $\phi_q = 45^\circ$ (solid red).
As before, the carrier concentration is $n = 5 \cdot 10^{14}$ m$^{-2}$,
corresponding to $E_F = 0.3 \cdot E_0$.
Region I marks again transitions with initial and final states in $E_{h1}$,
while in region II the final state is $E_{h2}$.
Not shown in the plot is the interband continuum with final states in $E_{l1}$ and $E_{l2}$, respectively.
Unlike GaAs, the interband and intraband parts merge at $\bar q \approx 0.6$.
The anisotropy in the plasmon spectrum is less pronounced than in GaAs.
The SPC, on the other hand, is only weakly direction dependent where
the main anisotropy arises in the intraband part.
Contrary to GaAs, the two modes enter the continuum at almost equally large momenta $\bar q \approx 0.45$.

\begin{figure}[b]
\includegraphics[width=\linewidth]{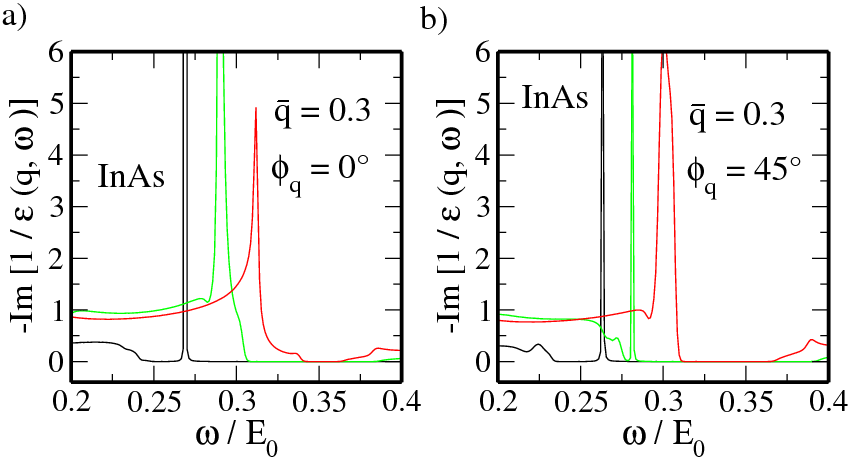}
\caption{(Color online) Energy loss function of InAs for fixed $\bar q = 0.3$
and various electric fields $E_z = 0$ (black), $1.8\cdot 10^7$~V/m (green) and $2.8\cdot 10^7$~V/m (red) for two different angle orientations $\phi_q = 0^\circ$ (a) and $45^\circ$ (b).
The hole density is set to $n=5\cdot 10^{14}$~m$^{-2}$ ($E_F = 0.3 \cdot E_0$).
}
\label{lossfunction_InAs_electricfield}
\end{figure}

\begin{figure}[tb]
\includegraphics[width=\linewidth]{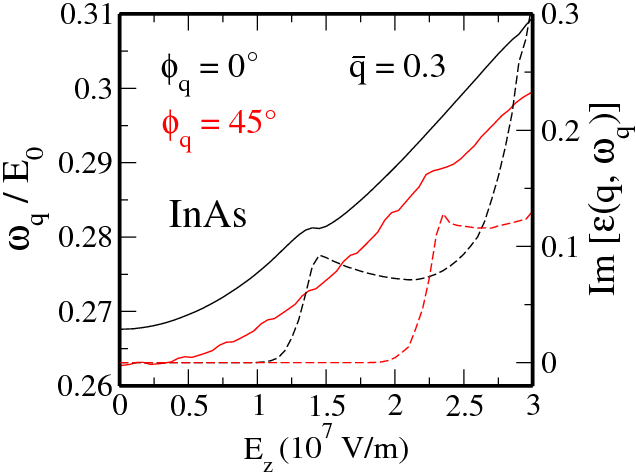}
\caption{(Color online) Electric field dependence of the plasmon energy $\omega_q$ (solid lines) and lifetime being proportional to the
imaginary part of the dielectric function $\Im{\varepsilon (q, \omega_q)}$ (dashed line) for fixed wave vector $\bar q = 0.3$
and two different angle orientations, $\phi_q = 0^\circ$ (black) and $\phi_q = 45^\circ$ (red).
The hole density is again set to $n=5\cdot 10^{14}$~m$^{-2}$ ($E_F = 0.3 \cdot E_0$).
}
\label{Plasmons_InAs_Efield_dependence}
\end{figure}

Up to now, spin-splitting of the energy bands is only caused by the Dresselhaus term.
The effect of an increasing electric field, pointing along the growth direction, which leads to a finite Rashba contribution in Eq.~(\ref{Hamiltonian_HR}), is demonstrated in Fig.~\ref{lossfunction_InAs_electricfield} for fixed momentum $\bar q = 0.3$ and
two different directions $\phi_q = 0^\circ$ and $\phi_q = 45^\circ$.
The $\delta$-like peak in the case without applied external electric field becomes broadened for large enough $E_z$,
because both the intra- and interband continua expand due to the spin-splitting.
If the interband SPC is sufficiently broad, it finally contains the point of the resonant energy.
The detailed field dependence of the plasmon energy and lifetime is shown in Fig.~\ref{Plasmons_InAs_Efield_dependence}.
The energy (solid line) is blueshifted once the electric field is enlarged whereat
the difference between the $E_z = 0$ and $E_z = 3.0 \cdot 10^7$ V/m result is about $15$ percent.

More important than the position of the peak is its width, which in turn is related to the imaginary part of the dielectric function
(dashed lines).
For $\phi_q = 0^\circ$ the mode remains undamped if $E_z  \lesssim 1.0 \cdot 10^7$ V/m.
For larger fields, the quasiparticles acquire a finite lifetime
as $\Im{\varepsilon(q,\omega_q)}$ is nonzero.
Qualitatively the same features can be seen for $\phi_q = 45^\circ$.
However, the important point is that the critical field at which the mode enters the SPC, $E_z \approx 1.9 \cdot 10^7$ V/m, is now much larger.
This in turn yields two possible applications. First, in a plasmon field-effect transistor, 
excitations in the source can be detected in the drain, depending on their lifetime
and thus on the strength of the applied electric field.\cite{Li_2008}
Second, changing the electric field in the range of $1.0 \cdot 10^7 \text{ V/m} < E_z < 1.9 \cdot 10^7 \text{ V/m}$ allows for
plasmon filtering since particular directions are damped while others are not.\cite{Badalyan_2009}

Finally, we want to remark that as in GaAs, additional solutions of the approximate solution Eq.~(\ref{Def_weakly_damped_Plasmons})
can be found in InAs.
However, as can be seen in Fig.~\ref{Plasmons_2nd_zero}(b), only one of these solutions corresponds to a genuine plasmonic mode
with a characteristic resonance in the energy loss function.

\section{Conclusions and outlook}
\label{sect:Conclusions}
We have investigated the dynamical dielectric functions of a GaAs and InAs quantum well, including spin-orbit interactions of 
the lowest order Dresselhaus and Rashba type, in the four-band Luttinger Hamiltonian
for the two lowest subbands.
For GaAs a pronounced anisotropy in the plasmon spectrum and in the single-particle continuum occurs.
Depending on the direction of the incident beam, plasmons are damped or long-lived.
This opens, in principle, the possibility of filtering plasmons with distinct orientations.\cite{Badalyan_2009, Li_2008}
However, the main problem in GaAs is the disability of controlling this feature due to the negligible Rashba contribution.

Here InAs seems to be more promising. In fact, it turns out that while long-wavelength plasmons in InAs
do not decay provided the Rashba contribution is small enough,
the lifetime of the plasmons can be modulated by changing the electric field.
Furthermore, we have shown that the critical field at which a finite damping occurs is direction dependent.
These features might be useful for applications such as a plasmon field-effect transistor\cite{Li_2008}
or a plasmon filter.\cite{Badalyan_2009, Li_2008}

Finally let us point out that
while this study is based on realistic standard 
band parameters,\cite{Winkler_Book} the strength of the spin-orbit coupling
contributions can even be enlarged beyond that values
by applying strain to the sample or using more advanced heterostructures
which might lead to an enhancement of the effects described in this work.

\acknowledgments
We thank T. Stauber for useful discussions.
This work was supported by Deutsche Forschungsgemeinschaft via Grant No.~SFB 689.



\begin{thebibliography}{99}

\bibitem{Zutic_2004}
I. Zutic, J. Fabian, and S. Das Sarma, Rev. Mod. Phys. \textbf{76}, 323 (2004).

\bibitem{Fabian_2007}
J. Fabian, A. Matos-Abiague, C. Ertler, P. Stano, and I. Zutic, Acta Phys. Slov. \textbf{57}, 565 (2007). 

\bibitem{Wu_2010}
M.~W. Wu, J.~H. Jiang, and M.~Q. Weng, Phys. Rep. \textbf{493}, 61 (2010).

\bibitem{Li_2008}
C. Li and X.~G. Wu, Appl. Phys. Let. \textbf{93}, 251501 (2008).

\bibitem{Badalyan_2009}
S.~M. Badalyan, A. Matos-Abiague, G. Vignale, and J. Fabian,
Phys. Rev. B {\bf 79}, 205305 (2009).

\bibitem{Kyrychenko_2009}
F.~V. Kyrychenko and C.~A. Ullrich, J. Phys.: Condens. Matter \textbf{21} 084202 (2009).

\bibitem{Guil} 
G. Giuliani and G. Vignale, \textit{Quantum Theory of the Electron Liquid} 
(Cambridge University Press, Cambridge, 2005).

\bibitem{Fetter}
A.~L. Fetter and J.~D. Walecka, \textit{Quantum Theory of Many-Particle Systems}
(Dover Publications, Mineola, 2003).

\bibitem{Badalyan_2010}
S.~M. Badalyan, A. Matos-Abiague, G. Vignale, and J. Fabian,
Phys. Rev. B {\bf 81}, 205314 (2010).

\bibitem{Pletyukhov_2007}
M. Pletyukhov and S. Konschuh, Eur. Phys. J. B \textbf{60}, 29 (2007).

\bibitem{Ullrich_2003}
C.~A. Ullrich and M.~E. Flatte, Phys. Rev. B {\bf 68}, 235310 (2003).

\bibitem{Kyrychenko_2011}
F.~V. Kyrychenko and C.~A. Ullrich, Phys. Rev. B \textbf{83}, 205206 (2011).

\bibitem{Schliemann_2010}
J. Schliemann, Phys. Rev. B {\bf 74}, 045214 (2006); \textit{ibid} {\bf 84}, 155201 (2011);
Europhys. Lett. {\bf 91}, 67004 (2010).

\bibitem{Kernreiter_2010}
T. Kernreiter, M. Governale, and U. Z\"ulicke, New J. Phys. {\bf 12}, 093002 (2010).

\bibitem{Cheng_2001}
S.~J. Cheng and R.~R. Gerhardts, Phys. Rev. B \textbf{63}, 035314 (2001).

\bibitem{Wunsch} 
B. Wunsch, T. Stauber, F. Sols, and F. Guinea, 
New J. Phys. \textbf{8}, 316 (2006).

\bibitem{Sarma} 
E.~H. Hwang and S. Das Sarma, Phys. Rev. B \textbf{75}, 205418 (2007).

\bibitem{Stauber_2010}
T. Stauber, J. Schliemann, and N.~M.~R. Peres,
Phys. Rev. B {\bf 81}, 085409 (2010).

\bibitem{Wang_2007_2}
X.-F. Wang and T. Chakraborty, Phys. Rev. \textbf{B} 75, 033408 (2007).

\bibitem{Pyat} 
P.~K. Pyatkovskiy, J. Phys.: Condens. Matter \textbf{21}, 025506 (2009).

\bibitem{Scholz_2011}
A. Scholz and J. Schliemann, Phys. Rev. B \textbf{83}, 235409 (2011).

\bibitem{Scholz_2012}
A. Scholz, T. Stauber, and J. Schliemann, Phys. Rev. \textbf{B} 86, 195424 (2012).

\bibitem{Luttinger_1956}
J.~M. Luttinger, Phys. Rev. \textbf{102}, 1030 (1956).

\bibitem{Dresselhaus_1955}
G. Dresselhaus, Phys. Rev. \textbf{100}, 580 (1955).

\bibitem{Rashba_1984}
Y. Bychkov and E.~I. Rashba, JETP Lett. \textbf{39}, 78 (1984).

\bibitem{Winkler_Book}
R. Winkler, \textit{Spin-Orbit Coupling Effects in Two-Dimensional Electron and Hole Systems} 
(Springer Tracts in Modern Physics, Berlin Heidelberg, 2003).

\bibitem{Badalyan_2008}
S.~M. Badalyan, C.~S. Kim, and G. Vignale, Phys. Rev. Lett. \textbf{100}, 016603 (2008).

\bibitem{Schueller_1994}
C. Sch\"uller, J. Kraus, G. Schaack, G. Weimann, and K. Panzlaff, Phys. Rev. B \textbf{50}, 18387 (1994).

\bibitem{Kane_1957}
E.~O. Kane, J. Phys. Chem. Solids \textbf{1}, 249 (1957). 

\end{thebibliography}
\end{document}